\documentclass{article}
\usepackage{spconf,amsmath,graphicx}

\usepackage{amsmath,cite,url}
\usepackage{graphicx}
\usepackage{color}

\usepackage{tikz}
\usepackage{multirow}
\usepackage{subcaption}
\usepackage{amssymb}
\usepackage[mathscr]{eucal}
\usepackage{amsmath,amssymb,amsfonts}
\usepackage{algorithmic}
\usepackage{graphicx}
\usepackage{textcomp}
\usepackage{xcolor}
\usepackage{wrapfig}

\usepackage{todonotes}

\newcommand{\drawmatr}[6]{
    \draw[black,fill=#6] (#1,#2,#3) -- ++(#4,0,0) -- ++(0,-#5,0) -- ++(-#4,0,0) -- cycle;
}

\title{Exploring single-song autoencoding schemes \\ for audio-based music structure analysis}%structural segmentation of music}
\name{Axel Marmoret, Jérémy E. Cohen\sthanks{This work is partly supported by ANR JCJC project LoRAiA ANR-20-CE23-0010.}, Frédéric Bimbot}
\address{Univ Rennes, Inria, CNRS, IRISA, France.}
\begin{document}
%\ninept
%
\maketitle
\begin{abstract}
The ability of deep neural networks to learn complex data relations and representations is established nowadays, but it generally relies on large sets of training data.
This work explores a ``piece-specific'' autoencoding scheme, in which a low-dimensional autoencoder is trained to learn a latent/compressed representation specific to a given song, which can then be used to infer the song structure. %\jc{We propose to model each song individually with an autoencoder learning similarities and differences between segments and scattering them in a latent space.}{remove?}
Such a model does not rely on supervision nor annotations, which are well-known to be tedious to collect and often ambiguous in Music Structure Analysis. %\jc{This work evaluates an elementary autoencoder on the RWC-Pop dataset, \jc{whose}{which} songs are represented with different features.}{remove?}
We report that the proposed unsupervised auto-encoding scheme achieves the level of performance of supervised state-of-the-art methods with 3 seconds tolerance when using a Log Mel spectrogram representation on the RWC-Pop dataset.
%\jc{We report results where some of them perform as well as the current best ``blind'' approaches, and}{We report that the proposed unsupervised auto-encoding scheme} achieves the level of performance of supervised state-of-the-art methods with 3 seconds tolerance \jc{}{when using a Mel-spectrogramm representation}{}.
%We evaluate several autoencoding variants on the RWC-Pop dataset and we report results where some of them perform as well as the current best ``blind'' approaches, and can achieve a level of performance comparable to supervised state-of-the-art methods in optimal oracle conditions.
\end{abstract}
\begin{keywords}
Autoencoders, Music Structure Analysis, Audio Signal Processing
\end{keywords}

\section{Introduction}\label{sec:introduction}

Music Structure Analysis (MSA) consists in segmenting a music piece in several distinct parts, which represent a mid-level description. Segmentation is usually based on criteria such as 
%These distinct parts are defined as internally coherent, and their retrieval generally consists in finding the most homogeneous ones or boundaries between two strongly dissimilar ones.
homogeneity, novelty, repetition and regularity~\cite{nieto2020segmentationreview}. In practice, MSA often relies on similarity between passages of a song summarized in an autosimilarity matrix~\cite{foote2000automatic, mcfee2014analyzing, nieto2013convex, marmoret2020uncovering}, in which each coefficient represents an estimation of the similarity between two musical fragments.

Recently, Deep Neural Networks (DNN) have lead to some high level of excitement in Music Information Retrieval (MIR) research, and notably in MSA~\cite{grill2015music, mccallum2019unsupervised}. In general, DNN approaches rely on large databases which make it possible to learn a large number of parameters, which in turn yields better performance than previously established machine learning approaches. This is the consequence of the ability of DNNs to learn complex nonlinear mappings through which musical objects can be well separated~\cite{humphrey2013feature}.

However, DNNs generally learn and exploit ``deep'' features stemming from multiple examples used in a training phase, and then evaluate the potential of learned features in a test phase, where no optimization is performed.

In this work, we consider a different DNN approach: we train a song-dependent autoencoder (AE), \textit{i.e.} an AE which is specifically trained to compress a given song. By doing so, we make the hypothesis that the ability of AE to learn compressed representations of musical objects occurring within a given song will provide a set of features that are useful to infer the structure of that song.

To study the relevance of this approach, this work presents a Convolutional Neural Network (CNN)-based technique for MSA, evaluated on the RWC-Pop and SALAMI datasets~\cite{goto2002rwc, smith2011design} in their audio form. Songs are represented with different features.
% and compare the representations obtained with different features on a same task, namely the structural segmentation of the RWC-Popular database~\cite{goto2002rwc} in its audio form. 

The rest of the article is structured as follows: Section 2 presents the formal concepts framing our work, Section 3 presents the details of our approach for music processing, Section 4 presents the segmentation process, and Section 5 reports on the experimental results on both datasets.

\section{Autoencoders}
\subsection{Basics of autoencoders}
\label{sec:ae_generalities}
Autoencoders are neural networks, which, by design, perform unsupervised dimensionality reduction. Throughout the years, AE have received increasing interest, notably due to their ability to extract salient latent representations without the need of large amount of annotations. Recently, AE also showed great results as a generation tool~\cite{engel2017neural}.%, roche2018autoencoders}.

%Concretely, given a generic entry $x \in \mathbb{R}^n$, an AE learns a nonlinear function $f$ with parameters $\theta$ (weights and biases of the network) such that $\hat{x} = f(x, \theta) \in \mathbb{R}^n$ reconstructs $x$ as faithfully as possible. This is achieved by minimizing a given loss function such as the Mean Square Error (MSE) between $x$ and $\hat{x}$ w.r.t. parameters $\theta$.
%\begin{equation}
%  \argmin_{\theta} \text{MSE}(x, \hat{x}) = \frac{1}{n} \sum_{i = 0}^{n - 1} (x_i - \hat{x}_i)^2~.
%\end{equation}
%Other metrics can be used, such as the KL-divergence, but we restrict this work to MSE.

An autoencoder is divided in two parts: an encoder, which compresses the input $x \in \mathbb{R}^n$ into a latent representation $z \in \mathbb{R}^m$ of smaller dimension (generally, $m \ll n$), and a decoder, which reconstructs $\hat{x}$ from $z$.

%A shallow encoder is constructed with one layer $W^e$, a bias $b^e$, and a nonlinear activation function $\sigma$, such that $z=\sigma(W^ex + b^e)$. The decoder follows as $\hat{x} = \sigma(W^dz + b^d)$

%``Deep'' autoencoders use the same formalism, but by stacking several layers ; that is, for an encoder with $l$ layers, $z = \sigma_l(W^e_l \sigma_{l-1}(W^e_{l-1} \sigma_{l-2}(... (W^e_1x + b^e_1)) + b^e_{l-1}) + b^e_l)$. In general, the decoder mimics the architecture of the encoder, that is, if the $i^{th}$ layer of the encoder $W^e_i$ has sizes $k \times k'$, then the $i^{th}$ layer of decoder $W^d_i$ has sizes $k' \times k$. We apply this strategy in what follows.

In this work, layers are of two types: fully-connected, or convolutional. %While fully-connected layers connect every neurons of two successive layers with independent weights and bias, which have to be treated as individual parameters, convolutional layers are made of small kernels which are convolved with parts of the input and share their weights. %This condition lowers the number of parameters to be learned.
Convolutional layers lead to impressive results in image processing due to their ability to discover local correlations (such as lines or edges), which turn to higher-order features with the depth of the network~\cite{lecun1998gradient}. 
%~\cite[Ch.~9]{Goodfellow2016}
While local correlations are less obvious in spectrogram processing~\cite{peeters2021deep}, Convolutional Neural Networks (CNN) still perform well in MIR tasks, such as MSA~\cite{grill2015music}.

\subsection{AE architecture}
The tested AE is a CNN. We use the nowadays quite standard ReLu function as the activation function, except in the latent layer because it could lead to null latent variables. %Hence, the latent layer is not composed of a nonlinear activation function.
The encoder is composed of five hidden layers: 2 convolutional/max-pooling blocks, followed by a fully-connected layer, controlling the size of the latent space. Convolutional kernels are of size 3x3, and the pooling is of size 2x2. Convolutional layers define respectively 4 and 16 feature maps.

\begin{figure}[htbp]
 \centerline{\includegraphics[width=\columnwidth]{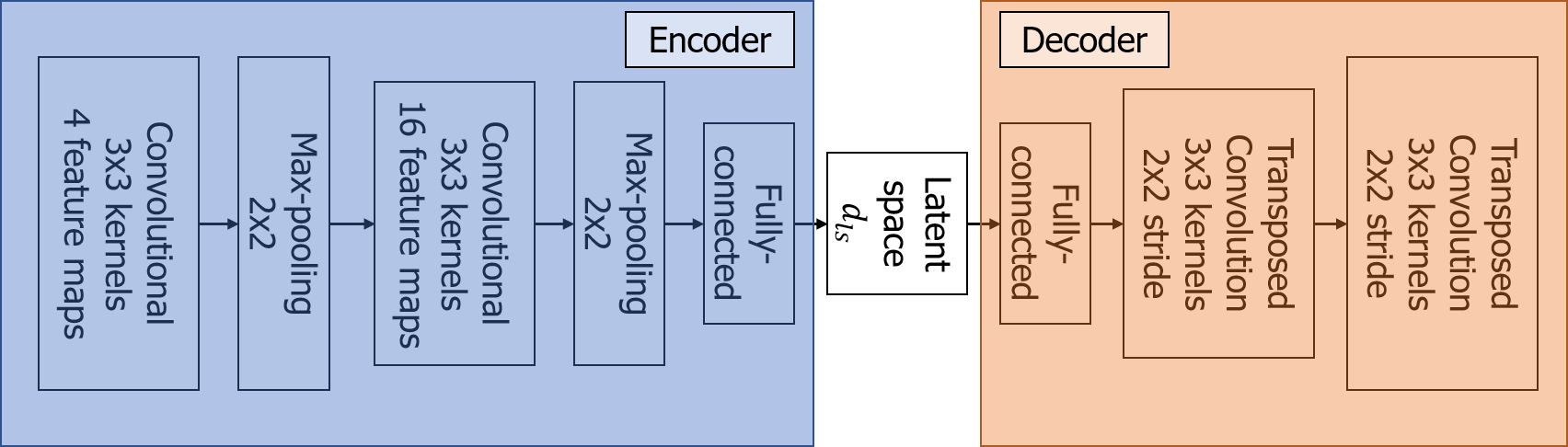}}
\setlength{\belowcaptionskip}{-20pt}
 \caption{Architecture of the autoencoder}
 \label{fig:convolutional}
\end{figure}

The decoder is composed of 3 hidden layers: a fully-connected layer (inverse of the previous one) and 2 ``transposed convolutional'' layers of size 3x3 and stride 2x2. A transposed convolution is similar to the convolution operation taken in the backward pass: an operation which takes one scalar as input and returns several scalars as output~\cite[Ch.~4]{dumoulin2016guide}. %~\cite[Ch.~9]{Goodfellow2016}.
Hence, it is well suited to inverse the convolution process. This network is represented in Figure~\ref{fig:convolutional}.

\section{AE for Music Processing}
\subsection{Motivations}
The underlying idea of this work is that music structure can be related to compression of information. Indeed, a common view of music structure is to consider structural segments as internally coherent passages, and automatic retrieval techniques generally focus on finding them by maximizing homogeneity and repetition and/or by setting boundaries between dissimilar segments, as points of high novelty. 
In the context of compressed representations, such as AEs, each passage is transformed in a vector of small dimension, compelling this representation to summarize the original content.
From this angle, similar passages are expected to be represented by similar representations, as they share underlying properties (such as coherence and redundancy), while dissimilar passages are bound to create strong discrepancies at their boundaries. Thus, we expect that compressed representations will enhance the original structure while reducing incidental signal-wise properties which do not participate to the structure.

\subsection{Barwise music processing}
Following our former work in~\cite{marmoret2020uncovering}, we process music as barwise spectrograms, with a fixed number of frames per bar.

%Conversely to fixed-size frame analysis, barwise computation reduces the impact of tempo in the information extracted to represent the local properties of the musical flow, as it is based on the metric, which is a more abstract musical notion to describe time. 
Conversely to fixed-size frame analysis, barwise computation guarantees that the information contained in each frame does not depend on the tempo, but on the metrical positions, which is a more abstract musical notion to describe time.
As a consequence, comparisons of bars are more reliable than comparisons of frames of arbitrarily fixed size
%two different bars should be more comparable than two frames (or sets of frames) of arbitrarily fixed size,
as it allows to cope with small variations of tempo.% between different bars.

In addition, pop music (\textit{i.e.} our case study) is generally quite regular at the bar level: repetitions occur rather clearly at the bar scale (or its multiples) and motivic patterns tend to develop within a limited number of bars, suggesting that frontiers mainly occur between bars.

Accordingly, this method relies on a consistent bar division of music. It also requires a powerful tool to detect bars, as otherwise errors could propagate and affect the performance. Results reported in~\cite{marmoret2020uncovering} tend to show that the madmom toolbox~\cite{madmom} is efficient in that respect on the RWC-Pop dataset. Nonetheless, this processing may hinder the retrieval of boundaries based on a change of tempo.%, as discussed in~\cite{vatolkin2021evolutionary}.

%The proposed approach relies on some consistent bar division of music, which is generally the case for contemporary western music, but is not a universal rule, and/or may turn out to be ambiguous.

Following this approach, each bar is represented with 96-equally spaced frames, selected from a high-resolution spectrogram (computed with a hop-length of 32 samples)~\cite{marmoret2020uncovering}. %This time-representation allows to cope with small variations of tempo between different bars, while preserving consistent bar-scale structure across bars.

\subsection{Features}
Music is represented with different features throughout our experiments, focusing on different aspects of music such as harmony or timbre.%\AxelNote{Axel: Abréger cette partie, juste le lister ?}\todo{Jeremy: Oui au minimum la liste à puce est accessoire.}

\textbf{Chromagram} A chromagram represents the time-frequency aspect of music as sequences of 12-row vectors, corresponding to the 12 semi-tones of the classical western music chromatic scale (C, C\#, ..., B), which is largely used in Pop music. Each row represents the weight of a semi-tone (and its octave counterparts) at a particular instant.
    
\textbf{Mel spectrogram} A Mel spectrogram corresponds to the STFT representation of a song, whose frequency bins are recast in the Mel scale. %These bands account for the exponential spread of frequencies throughout the octaves. 
Mel spectrograms are dimensioned following the work of~\cite{grill2015music} (80 filters, from 80Hz to 16kHz). STFT are computed as power spectrograms.

\textbf{Log Mel spectrogram} A Log Mel spectrogram corresponds to the logarithmic values of the precedent spectrogram. This representation accounts for the exponential decay of frequency power.
    
\textbf{MFCC spectrogram} Mel-Frequency Cepstral Coefficients (MFCCs) are timbre-related coefficients, obtained by a discrete cosine transform of the Log Mel spectrogram. This spectrogram contains 32 coefficients, following~\cite{mcfee2014learning}.

%In summary, the AEs learn features on barwise spectrograms $\times$ 96 time indexes. Each song is then represented as $B$ matrices, with $B$ being its number of bars.

%Due to the variety of music, chroma information and 4/4 metric are hypotheses which may not be adequate for all music genres. Both of these limits can be adapted to music specificities:
%\begin{itemize}
%\item Different tonal/modal systems can be coped with using different representations (CQT, STFT, ...).
%\item To account for different metrics, a metric estimator can be used along with bar segmentation, to inform the neural network of the accurate kernel size. As a first attempt, we did not follow that track.
%\end{itemize}

\section{Structural segmentation}
The ability of a single-song AE to separate and group bars is evaluated on the MSA task, as presented in~\cite{paulus2010state}.

Thereby, for a given song in a given representation, we compute latent representations ${z_b}$ for all bars $1 \leq b \leq B$ of the song. These latent representations are at the heart of the segmentation strategy, via their autosimilarity matrix.

\begin{figure}[ht] 
\vspace{-6pt}
 \includegraphics[width=\columnwidth]{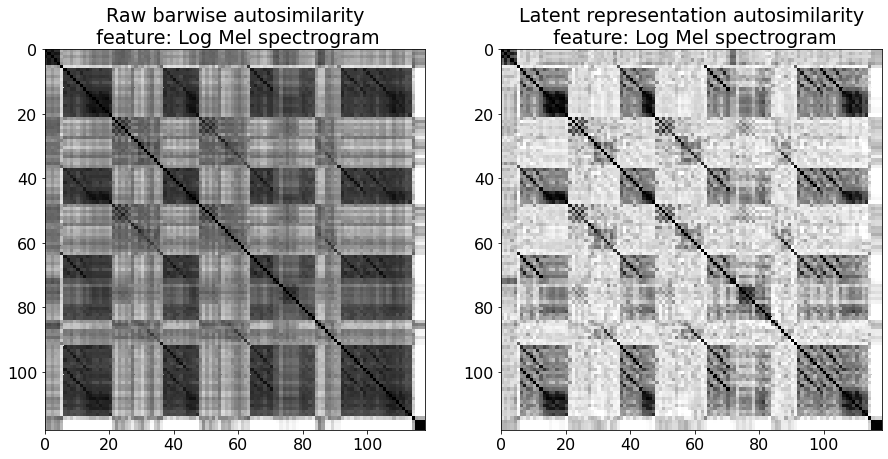}
 \vspace{-13pt}
\setlength{\belowcaptionskip}{-5pt}
 \caption{Barwise autosimilarity of the raw Log Mel spectrogram (left) and of the latent representations (right), computed on the song ``Pop01'' from RWC-Pop.}% Grey lines represent the annotation.}
 \label{fig:autosimilarity}
\end{figure}

Denoting $Z \in \mathbb{R}^{d_{ls} \times B}$ the matrix resulting from the concatenation of all $d_{ls}$-dimensional $z_b$ vectors, its autosimilarity is defined as $Z^T Z$, \textit{i.e.} the $B \times B$ matrix of the dot products between every $z_b$. Examples of autosimilarities are shown in Figure~\ref{fig:autosimilarity}.

Following the dynamic programming algorithm inspired from~\cite{sargent2016estimating} and presented in~\cite{marmoret2020uncovering}, the segmentation is computed as the global maximum of the sum of all individual segments costs. The cost of each segment is obtained by applying a sliding convolving kernel on the diagonal of the autosimilarity matrix. This convolving kernel is a square matrix, the size of which is that of the potential segment.

Contrary to the novelty kernel of Foote~\cite{foote2000automatic}, largely described in the literature~\cite{nieto2020segmentationreview}, the proposed kernel aims at framing square blocks of high similarity, occurring when consecutive bars are similar. In that sense, while Foote's kernel focuses on novelty, ours favors homogeneity. Still, the dynamic programming algorithm allows to find a trade-off between maximizing homogeneity of individual blocks and novelty across them as a whole. When the diagonal of the autosimilarity is structured in several self-similar blocks, the algorithm frames and partitions these blocks.

\begin{wrapfigure}{r}{0.35\columnwidth}
\setlength{\columnsep}{-2pt}%
  \vspace{-21pt}
  \begin{center}
    \begin{tikzpicture}[scale=2.05]
    \drawmatr{0}{0}{0}{1}{1}{gray}
    \drawmatr{0}{-0.1}{0}{0.1}{0.4}{black}
    \drawmatr{0}{0}{0}{0.1}{0.1}{white}
    \drawmatr{0.1}{-0.1}{0}{0.1}{0.1}{white}
    \drawmatr{0.2}{-0.2}{0}{0.1}{0.1}{white}
    \drawmatr{0.3}{-0.3}{0}{0.1}{0.1}{white}
    \drawmatr{0.4}{-0.4}{0}{0.1}{0.1}{white}
    \drawmatr{0.5}{-0.5}{0}{0.1}{0.1}{white}
    \drawmatr{0.6}{-0.6}{0}{0.1}{0.1}{white}
    \drawmatr{0.7}{-0.7}{0}{0.1}{0.1}{white}
    \drawmatr{0.8}{-0.8}{0}{0.1}{0.1}{white}
    \drawmatr{0.9}{-0.9}{0}{0.1}{0.1}{white}
    \drawmatr{0.1}{-0.2}{0}{0.1}{0.4}{black}
    \drawmatr{0.2}{-0.3}{0}{0.1}{0.4}{black}
    \drawmatr{0.3}{-0.4}{0}{0.1}{0.4}{black}
    \drawmatr{0.4}{-0.5}{0}{0.1}{0.4}{black}
    \drawmatr{0.5}{-0.6}{0}{0.1}{0.4}{black}
    \drawmatr{0.6}{-0.7}{0}{0.1}{0.3}{black}
    \drawmatr{0.7}{-0.8}{0}{0.1}{0.2}{black}
    \drawmatr{0.8}{-0.9}{0}{0.1}{0.1}{black}
    \drawmatr{0.1}{0}{0}{0.4}{0.1}{black}
    \drawmatr{0.2}{-0.1}{0}{0.4}{0.1}{black}
    \drawmatr{0.3}{-0.2}{0}{0.4}{0.1}{black}
    \drawmatr{0.4}{-0.3}{0}{0.4}{0.1}{black}
    \drawmatr{0.5}{-0.4}{0}{0.4}{0.1}{black}
    \drawmatr{0.6}{-0.5}{0}{0.4}{0.1}{black}
    \drawmatr{0.7}{-0.6}{0}{0.3}{0.1}{black}
    \drawmatr{0.8}{-0.7}{0}{0.2}{0.1}{black}
    \drawmatr{0.9}{-0.8}{0}{0.1}{0.1}{black}

    \drawmatr{1.1}{-0.2}{0}{0.1}{0.1}{black}
    \node at (1.3,-0.25) {2};
    \drawmatr{1.1}{-0.4}{0}{0.1}{0.1}{gray}
    \node at (1.3,-0.45) {1};    \drawmatr{1.1}{-0.6}{0}{0.1}{0.1}{white}
    \node at (1.3,-0.65) {0};

\end{tikzpicture}
\end{center}
  \setlength{\abovecaptionskip}{-6pt}%
  \setlength{\belowcaptionskip}{-13pt}%
  \caption{Kernel of size 10}
  %\vspace{-10pt}
\end{wrapfigure}
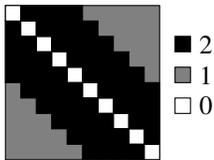

%Concretely, the proposed kernel is equal to 0 on the diagonal, 2 on the lower and upper 4 subdiagonals, and 1 everywhere else. 
The diagonal is equal to zero, so as to disregard perfect self-similarity of each bar with itself (normalized to 1).
The other elements of this kernel are designed so as to emphasize the short-term similarity in the 4 contiguous bars, and still catch longer-term similarity for long segments. This kernel performs best in comparative experiments\footnote{https://gitlab.inria.fr/amarmore/musicntd/-/blob/v0.2.0/Notebooks/5\%20-\%20Different\%20kernels.ipynb} of~\cite{marmoret2020uncovering}.

This raw convolutional cost is combined with a(n ir-)regularity cost, favoring frequent segments' sizes~\cite{sargent2016estimating}, adapted as in~\cite{marmoret2020uncovering}.

%To account for regularity constraints as in~\cite{sargent2016estimating}, the present algorithm adds a (ir-)regularity penalty $p(n)$ to the plain convolution score, which is a function of the size $n$ (in bars) of the segment. This function $p(n)$ is equal to 0 if $n = 8$, $\frac{1}{4}$ if $n \equiv 0 \pmod 4$, $\frac{1}{2}$  if $n \equiv 0 \pmod 2$, and finally 1 otherwise. This penalty function is musically motivated for pop music, as segments are more likely to be of size 8 bars (especially in RWC-Pop~\cite{sargent2016estimating}), or multiple of 4 bars, than of odd bar size.

%Finally, for all possible segments $b_1, b_2$ in the song, the algorithm computes a score $s_{b_1,b_2} = \frac{c_{b_1,b_2}}{c_{k8}^{max}} - p(n)$, where $c_{b_1,b_2}$ is the convolution cost, and where $c_{k8}^{max}$ is the highest convolution score on this autosimilarity with a kernel of size 8, used for normalization purposes.

\section{Experiments}
This segmentation pipeline is studied on the RWC-Pop dataset, which consists in 100 Pop songs of high recording quality~\cite{goto2002rwc}, along with the MIREX10 annotations~\cite{bimbot2014semiotic}. The best autoencoding method is also evaluated on the SALAMI dataset~\cite{smith2011design}. We focus on boundary retrieval and ignore segment labelling.

Boundaries are evaluated using the hit-rate metric, which considers a boundary valid if it is approximately equal to an annotation, within a fixed tolerance window. Consistently with MIREX standards~\cite{MIREXsite}, tolerances are equal to 0.5s and 3s. The hit-rate is then expressed in terms of Precision, Recall, and F-measure.

\subsection{Related work}

\begin{table*}[ht]
\caption{Results of best-performing AEs and state-of-the-art on the RWC-Pop dataset. (*Results: 2015 MIREX contest~\cite{MIREXsite}.)}
\vspace{-10pt}
\setlength{\abovecaptionskip}{-10pt}%
\setlength{\belowcaptionskip}{-20pt}%
 \begin{center}
\begin{tabular}{ll|c|c|c|c|c|c|}
\cline{3-8} &        & $P_{0.5}$ & $R_{0.5}$ & $F_{0.5}$ & $P_{3}$ & $R_{3}$ & $F_{3}$ \\
\hline
\multicolumn{1}{|l|}{\multirow{4}{*}{\begin{tabular}[c]{@{}l@{}}Latent\\ representations\\ autosimilarity\end{tabular}}}           & \multicolumn{1}{l|}{Chromagram}  & \multicolumn{1}{l|}{51.8\%}    & \multicolumn{1}{l|}{54\%}    & \multicolumn{1}{l|}{52.3\%}    & \multicolumn{1}{l|}{70.5\%}  & \multicolumn{1}{l|}{73.3\%}  & \multicolumn{1}{l|}{71.2\%}  \\ \cline{2-8} 
\multicolumn{1}{|l|}{}& \multicolumn{1}{l|}{Mel Spectrogram}   & \multicolumn{1}{l|}{52.1\%}          & \multicolumn{1}{l|}{55.3\%}          & \multicolumn{1}{l|}{53.3\%}          & \multicolumn{1}{l|}{73.1\%}        & \multicolumn{1}{l|}{77.6\%}        & \multicolumn{1}{l|}{74.7\%}        \\ \cline{2-8} 
\multicolumn{1}{|l|}{}  & \multicolumn{1}{l|}{Log Mel Spectrogram}  & \multicolumn{1}{l|}{59.5\%}    & \multicolumn{1}{l|}{61.3\%}    & \multicolumn{1}{l|}{59.9\%}    & \multicolumn{1}{l|}{79.1\%}  & \multicolumn{1}{l|}{\textbf{81.7\%}}  & \multicolumn{1}{l|}{\textbf{79.9\%}}  \\ \cline{2-8} 
\multicolumn{1}{|l|}{}  & \multicolumn{1}{l|}{MFCC} & \multicolumn{1}{l|}{54.3\%}    & \multicolumn{1}{l|}{54.4\%}    & \multicolumn{1}{l|}{54\%}    & \multicolumn{1}{l|}{76.6\%}  & \multicolumn{1}{l|}{76.4\%}  & \multicolumn{1}{l|}{76\%}  \\ \hline \hline

\multicolumn{1}{|l|}{\multirow{4}{*}{\begin{tabular}[c]{@{}l@{}}Barwise\\ feature\\ representation \\autosimilarity\end{tabular}}}           & \multicolumn{1}{l|}{Chromagram}  & \multicolumn{1}{l|}{39.6\%}    & \multicolumn{1}{l|}{38.4\%}    & \multicolumn{1}{l|}{38.6\%}    & \multicolumn{1}{l|}{63.8\%}  & \multicolumn{1}{l|}{61.2\%}  & \multicolumn{1}{l|}{62\%}  \\ \cline{2-8} 
\multicolumn{1}{|l|}{}  & \multicolumn{1}{l|}{Mel Spectrogram}   & \multicolumn{1}{l|}{46.4\%}          & \multicolumn{1}{l|}{47\%}          & \multicolumn{1}{l|}{46.3\%}          & \multicolumn{1}{l|}{70.9\%}        & \multicolumn{1}{l|}{71.3\%}        & \multicolumn{1}{l|}{70.6\%}        \\ \cline{2-8} 
\multicolumn{1}{|l|}{} & \multicolumn{1}{l|}{Log Mel Spectrogram}  & \multicolumn{1}{l|}{46.5\%}    & \multicolumn{1}{l|}{46.1\%}    & \multicolumn{1}{l|}{46\%}    & \multicolumn{1}{l|}{72.2\%}  & \multicolumn{1}{l|}{71\%}  & \multicolumn{1}{l|}{71.1\%}  \\ \cline{2-8} 
\multicolumn{1}{|l|}{}  & \multicolumn{1}{l|}{MFCC} & \multicolumn{1}{l|}{41.9\%}    & \multicolumn{1}{l|}{40.6\%}    & \multicolumn{1}{l|}{40.9\%}    & \multicolumn{1}{l|}{68.1\%}  & \multicolumn{1}{l|}{65.6\%}  & \multicolumn{1}{l|}{66.4\%}  \\ \hline \hline

\multicolumn{1}{|l|}{\multirow{4}{*}{\begin{tabular}[c]{@{}l@{}}State-of-the-art\end{tabular}}} & \multicolumn{1}{l|}{Foote~\cite{foote2000automatic}} & \multicolumn{1}{l|}{42,0\%}   &  \multicolumn{1}{l|}{30,0\%}      & \multicolumn{1}{l|}{34,5\%}    & \multicolumn{1}{l|}{67,1\%}  & \multicolumn{1}{l|}{47,7\%}  & \multicolumn{1}{l|}{55,0\%}    \\ \cline{2-8} 
\multicolumn{1}{|l|}{} & \multicolumn{1}{l|}{Spectral clustering~\cite{mcfee2014analyzing}} & \multicolumn{1}{l|}{49.2\%}    & \multicolumn{1}{l|}{45\%}      & \multicolumn{1}{l|}{45\%}      & \multicolumn{1}{l|}{65.5\%}  & \multicolumn{1}{l|}{60.6\%}  & \multicolumn{1}{l|}{60.3\%}  \\ \cline{2-8} 
\multicolumn{1}{|l|}{}  & \multicolumn{1}{l|}{NTD~\cite{marmoret2020uncovering}}  & \multicolumn{1}{l|}{58.4\%}    & \multicolumn{1}{l|}{60.7\%}    & \multicolumn{1}{l|}{59.0\%}      & \multicolumn{1}{l|}{72.5\%}  & \multicolumn{1}{l|}{75.3\%}  & \multicolumn{1}{l|}{73.2\%}  \\  \cline{2-8} 
\multicolumn{1}{|l|}{} & \multicolumn{1}{l|}{Supervised CNN~\cite{grill2015music}*} &  \multicolumn{1}{l|}{\textbf{80.4\%}}    & \multicolumn{1}{l|}{\textbf{62.7\%}}    & \multicolumn{1}{l|}{\textbf{69.7\%}}    & \multicolumn{1}{l|}{\textbf{91.9\%}}  & \multicolumn{1}{l|}{71.1\%}  & \multicolumn{1}{l|}{79.3\%}  \\ \hline

\end{tabular}
\end{center}
\label{table:table_cross_val}
\vspace{-20pt}
\end{table*}

Related work mainly splits in two categories: blind and learning-based techniques.

Blind segmentation techniques, such as this work, do not use training datasets and generally focus on autosimilarity matrices. The seminal paper in this area is that by Foote~\cite{foote2000automatic}, based on a novelty kernel, which detects boundaries as points of strong dissimilarity between the near past and future of a given instant. The strong point of this algorithm is its good performance given its relative simplicity.

Later on, McFee and Ellis made use of spectral clustering as a segmentation method by interpreting the autosimilarity as a graph and clustering principally connected vertices as segments~\cite{mcfee2014analyzing}. This work performed best among blind segmentation techniques in the last structural segmentation MIREX campaign in 2016~\cite{MIREXsite}. Recently, we used tensor decomposition (Nonnegative Tucker Decomposition, NTD) as a way to describe music as barwise patterns, which then served as features for the computation of an autosimilarity matrix~\cite{marmoret2020uncovering}.

Still, to the best of the authors' knowledge, the current state-of-the-art approach for the task of RWC-Pop dataset structural segmentation is a supervised CNN developed by Grill and Schlüter~\cite{grill2015music}. %, which obtained the best results in the 2015 MIREX campaign~\cite{MIREXsite}.
This technique also uses signal self-similarity as input, but in the form of lag matrices, and in conjunction with Log Mel spectrograms. Notably, this technique is trained on the SALAMI dataset only.

We compared the song-dependent autoencoders described in the present work with the aforementioned techniques. Results for~\cite{foote2000automatic} and~\cite{mcfee2014analyzing} were computed with the MSAF toolbox~\cite{nieto2016systematic}, and boundaries were aligned to the closest downbeat (as in~\cite{marmoret2020uncovering}). Results for~\cite{grill2015music} were taken from the 2015 MIREX campaign~\cite{MIREXsite}.

%\begin{figure}
%  \begin{subfigure}[ht]{0.8\columnwidth}
% \centering
%  \includegraphics[width=\columnwidth]{figs/res_0_5.png}
%  \caption{0.5 seconds tolerance results}
%\end{subfigure}
%\quad
%  \begin{subfigure}[ht]{0.8\columnwidth}
%      \includegraphics[width=\columnwidth]{figs/res_3.png}
%         \setlength{\abovecaptionskip}{-1pt}%
%      \caption{3 seconds tolerance results}
%  \end{subfigure}
% \setlength{\belowcaptionskip}{-12pt}%
% \caption{Results of the different networks when $d_{ls} = 32$}
%\end{figure}

\subsection{Practical considerations}

Networks are developed with Pytorch 1.8.0~\cite{NEURIPS2019_9015}. They are optimized with the Adam optimizer~\cite{kingma2014adam}, with a learning rate of 0.001, divided by 10 when the loss function reaches a plateau (20 iterations without improvement) until 1e-5. The optimization stops if no progress is made during 100 consecutive epochs, or after a total of 1000 epochs. Each song is processed in 8-size mini-batches. All networks are initialized with the uniform distribution defined in~\cite{he2015delving}, also known as ``kaiming'' initialization. Bars were estimated with madmom~\cite{madmom}, and spectrograms computed with librosa~\cite{mcfeeLibrosa}. All segmentation scores were computed with mir\_eval~\cite{raffel2014mireval}.

The entire code for this work is open-source, and contains experimental notebooks for reproducibility\footnote{https://gitlab.inria.fr/amarmore/musicae/-/tree/v0.1.0}. Performing 1000 epochs for a song takes between 1.5 minute (for chromagrams) and 5 minutes (for Mel/Log Mel spectrograms).

\begin{figure}[t]
 \includegraphics[width=0.8\columnwidth]{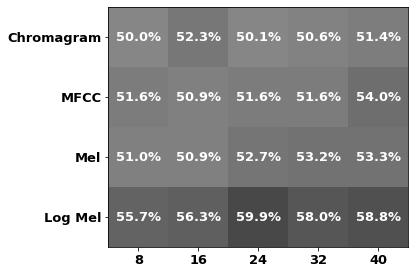}
\vspace{-6pt}

\setlength{\belowcaptionskip}{-8pt}
 \caption{F-measure with 0.5 seconds tolerance for AE with different features and latent space dimensions $d_{ls}$.}
 \label{fig:latent_dim}
 %\vspace{-5pt}
\end{figure}

\subsection{Results}

Figure~\ref{fig:latent_dim} presents results on the RWC-Pop dataset, when varying the dimension of the latent space $d_{ls}$. The Log Mel spectrogram clearly outperfoms the other features in this task, and achieves the state-of-the-art level of performance with 3-seconds tolerance, as presented in Table~\ref{table:table_cross_val}.

Table~\ref{table:res_salami} presents the results of autoencoding this Log Mel representation on the SALAMI dataset~\cite{smith2011design}. More precisely, we used the test subset of the 2015 MIREX competition, using the upper-level annotations, and keeping for each song the annotation resulting in the best $F_{3}$ among the two available. The latent space dimension was trained on the rest of the dataset, resulting in $d_{ls} = 32$. Even though they do not reach the state of the art level, results on SALAMI still appear competitive with a tolerance of 3 seconds.

Results on both datasets are still improvable with 0.5s tolerance window. A wrong estimation at 0.5s can be due to an incorrect frontier estimation, but also to an incorrect bar estimation. This could be particularly the case for the SALAMI dataset, where music is available in very different genres and recording conditions, such as live recordings.

\begin{table}[t]
\caption{Results on the SALAMI dataset (test subset of MIREX 2015), for the Log Mel spectrogram. Note: this subset contains song on which the CNN has been trained~\cite{grill2015music}.}
\vspace{-15pt}
\begin{center}
\begin{tabular}{l|c|c|}
\cline{2-3}
                                                 & $F_{0.5}$ & $F_3$ \\ \hline
\multicolumn{1}{|l|}{Latent representation autosimilarity}                & 36.6\%     & 59.3\%   \\ \hline
\multicolumn{1}{|l|}{Barwise feature autosimilarity} & 32\%     & 53.9\%   \\ \hline
\multicolumn{1}{|l|}{CNN~\cite{grill2015music} (taken from MIREX~\cite{MIREXsite})}  & 54.1\%     & 62.3\%   \\ \hline
\end{tabular}
\end{center}
\label{table:res_salami}
\vspace{-20pt}
\end{table}

\section{Conclusion}
Unsupervised AEs appear as competitive schemes for Music Structure Analysis. %Results suggest that sparsity leads to disentangled compressed representations which provide relevant features for structure retrieval. 
In our experiments, the Log Mel feature outperforms the other ones, and reaches the state-of-the-art at 3 seconds tolerance on the RWC-Pop music dataset.

Future work will focus on improving the proposed paradigm using strategies such as transfer learning from a song-independent AE or exploring various network architectures. Finally, while the convolutional dynamic programming algorithm is competitive, it is unstable to small variations in autosimilarities and should be made more robust.

Altogether, we believe that these first results pave the way to an interesting paradigm using single-song AEs for the description of structural elements in music.
%could focus on the robustness of the proposed approach with a sparsity-fitting heuristic. Another track 

\vfill\pagebreak

% References should be produced using the bibtex program from suitable
% BiBTeX files (here: strings, refs, manuals). The IEEEbib.bst bibliography
% style file from IEEE produces unsorted bibliography list.
% -------------------------------------------------------------------------
\bibliographystyle{IEEEbib}
\bibliography{biblio}

\end{document}